\documentstyle[12pt]{article}
\def\BorL{b }	
\makeatletter
\newif\ifpr@pstyle \pr@pstylefalse
\newif\ifnons@qeq  \nons@qeqfalse
\def\bigpage{
	\setlength{\topmargin}{-.5in}
	\setlength{\oddsidemargin}{.5pc}
	\setlength{\evensidemargin}{.5pc}
	\setlength{\textwidth}{35pc}
	\setlength{\textheight}{52pc}
	\setlength{\parskip}{6pt plus 2pt minus 1pt}
	\newlength{\paperbaselineskip}
	\setlength{\paperbaselineskip}{20pt plus 0.2pt minus 0.1pt}
	\def\@oddfoot{\hfil -- \thepage~--\hfil}
	\let\@evenfoot\@oddfoot
        \def\thesection{\arabic{section}.}
        \def\thesubsection{\thesection\arabic{subsection}}
        \def\@ourappendix{\par\setcounter{section}{0}
                      \setcounter{subsection}{0}
                      \def\thesection{\Alph{section}.}
                      \ifnons@qeq
                      \def\theequation{\Alph{section}.\arabic{equation}}\fi}
        \def\appendix{\@ourappendix}
        \def\section{\@startsection {section}{1}%
            {\z@}{5ex plus .2ex minus .4ex}%
            {1.5ex plus.4ex minus .1ex}%
            {\centering\ifpr@pstyle\else\reset@font\large\fi\bf}}
        \def\subsection{\@startsection{subsection}%
            {2}{\z@}{3.25ex plus .4ex minus .4ex}%
            {1ex plus .2ex}{\bf}}
}
\bigpage
\newfont{\fourteencp}{cmcsc10 scaled\magstep2}
\newfont{\titlefont}{cmbx10 scaled\magstep2}
\newfont{\authorfont}{cmcsc10 scaled\magstep1}
\newfont{\fourteenmib}{cmmib10 scaled\magstep2}
	\skewchar\fourteenmib='177
\newfont{\elevenmib}{cmmib10 scaled\magstephalf}
	\skewchar\elevenmib='177
\makeatletter
\newcommand\nonsequentialeqnum{
        \nons@qeqtrue
	\@addtoreset{equation}{section}
	\def\theequation{\arabic{section}.\arabic{equation}}}
\newif\ifp@bblock  \p@bblocktrue
\newcommand\nopubblock{\p@bblockfalse}
\newcommand\topspace{\hrule height 0pt depth 0pt \vskip}
\newcommand\p@bblock{\begingroup \tabskip=\hsize minus \hsize
	\baselineskip=1.5\ht\strutbox \topspace-2\baselineskip
	\halign to\hsize{\strut ##\hfil\tabskip=0pt\crcr
	\the\Pubnum\crcr\the\date\crcr}\endgroup}
\newcommand\YITPmark{\hbox{
        \ifpr@pstyle\twelvemib\else\fourteenmib\fi YITP\hskip0.2cm
        \ifpr@pstyle\tenmib\else\elevenmib\fi
           Uji\hskip0.15cm Research\hskip0.15cm Center\hfill}}
\newtoks\date
\newtoks\Pubnum
\newtoks\pubnum
\Pubnum={YITP/U-\the\pubnum}
\date={\today}
\newcommand{\frontpageskip}{\vspace{12pt plus .5fil minus 2pt}}
\def\@authoraddress{} \def\@title{}
\def\title#1{\gdef\@title{\frontpageskip
	\begin{center}{\titlefont #1}\end{center}\par}}
\def\@author#1{\frontpageskip\par\begin{center}{\authorfont #1}
	\end{center}
	\nobreak}
\def\author#1{\expandafter\def\expandafter\@authoraddress\expandafter
    {\@authoraddress{\@author{#1}}}}
\def\andauthor#1{\expandafter\def\expandafter\@authoraddress\expandafter
    {\@authoraddress{\frontpageskip\centerline{and}\@author{#1}}}}
\def\authors#1{\expandafter\def\expandafter\@authoraddress\expandafter
    {\@authoraddress{\frontpageskip\noindent #1}}}
\def\@address#1{\par\begin{center}{\sl #1}\end{center}\par}
\def\address#1{\expandafter\def\expandafter\@authoraddress\expandafter
    {\@authoraddress{\@address{#1}}}}
\def\andaddress#1{\expandafter\def\expandafter%
    \@authoraddress\expandafter
    {\@authoraddress{\par\centerline{\sl and}\@address{#1}}}}
\renewcommand{\thanks}[1]{\footnote{#1}}
\def\maketitle{\par
  \begingroup
       \def\thefootnote{\fnsymbol{footnote}}
	\thispagestyle{empty}
        \baselineskip=\paperbaselineskip
	\@maketitle
	\endgroup
	\setcounter{footnote}{0}
	\let\maketitle\relax \let\@maketitle\relax
	\let\@thanks\relax \let\@title\relax
	\let\@title\relax \let\@authoraddress\relax
	\let\thanks\relax}
\def\@maketitle{%
        \ifpr@pstyle\vspace{-1.0cm}\else\vspace{-1.7cm}\fi
	\YITPmark\vskip0.6cm
	\ifp@bblock\p@bblock \else\hrule height 0pt \relax \fi
	\@title
	\@authoraddress
	}
\renewcommand{\abstract}{\par\frontpageskip\centerline{
             \ifpr@pstyle\twelvecp\else\fourteencp\fi Abstract}
	\vspace{8pt plus 3pt minus 3pt}}
\makeatother
\def\bigmode{b }
\ifx\BorL\undefined\read-1 to\BorL\fi
\makeatletter
\def\doublepage{
        \twocolumn
        
        \pr@pstyletrue
        \sloppy
        \flushbottom
        \setlength{\topmargin}{-0.95in}
        \setlength{\headsep}{20pt}
        \setlength{\headheight}{10pt}
        \hoffset=-0.35in
        \leftmargini 2em
        \leftmarginv .5em
        \leftmarginvi .5em
        \marginparwidth 48pt
        \marginparsep 10pt
        \setlength{\columnsep}{0.7truein}
        \setlength{\textwidth}{10.5truein}
        \setlength{\textheight}{7truein}
        \setlength{\oddsidemargin}{0.0truein}
        \setlength{\evensidemargin}{0.0truein}
        \multiply\paperbaselineskip by 4
                   \divide\paperbaselineskip by 5
        \multiply\footskip by 4 \divide\footskip by 5
        \setlength{\parskip}{4pt plus 1.5pt minus 1pt}
        \newlength{\halfwidth}
        \halfwidth=\textwidth\advance\halfwidth by -\columnsep
                         \divide\halfwidth by 2
        \newfont{\twelvemib}{cmmib10 scaled\magstep1}
                 \skewchar\twelvemib='177
        \newfont{\tenmib}{cmmib10}
                 \skewchar\tenmib='177
        \newfont{\twelvecp}{cmcsc10 scaled\magstep1}
        \def\pagebox{\hbox to \halfwidth{\hfil  -- \thepage~--\hfil}}
        \def\@oddfoot{\pagebox\hfil\addtocounter{page}{1}\pagebox}
        \let\@evenfoot\@oddfoot
        \def\ps@empty{\let\@mkboth\@gobbletwo\let\@oddhead\@empty
               \def\@oddfoot{\hbox to \halfwidth{\hfil ~~~~~~~}\hfil
               \addtocounter{page}{1}\pagebox}
                \let\@evenhead\@empty\let\@evenfoot\@oddfoot}
        \def\appendix{\@ourappendix}
        \def\section{\@startsection {section}{1}%
            {\z@}{5ex plus .2ex minus .4ex}%
            {1.5ex plus.4ex minus .1ex}%
            {\centering\ifpr@pstyle\else\reset@font\large\fi\bf}}
        \def\subsection{\@startsection{subsection}%
            {2}{\z@}{3.25ex plus .4ex minus .4ex}%
            {1ex plus .2ex}{\bf}}
}
\ifx\BorL\bigmode
\else
        \input art10.sty
        \doublepage
\fi
\makeatother
\makeatletter
\newif\ifepsfloaded
\newif\iffigureexists

\ifeof 1 \epsfloadedfalse \else \epsfloadedtrue \fi
\ifepsfloaded \input epsf \fi

\def\checkex#1 {\relax
    \openin 1 #1
    \ifeof 1 \figureexistsfalse
    \else \figureexiststrue
    \fi \closein 1 }
\def\figinsertraw#1#2{
   \ifepsfloaded
       \checkex #1
       \iffigureexists
           \immediate\write16{(#1)}
           #2
       \else
           \immediate\write16{(#1 NOT FOUND!)}
           \vbox to 2in{\hbox to 2in {\hss} \vss}
       \fi
   \else
       \immediate\write16{(NOT inputting #1; no epsf.tex)}
       \vbox to 2in{\hbox to 2in {\hss} \vss}
   \fi}
\newcommand{\reduceland}[2]{\dimen@=#1
     \ifpr@pstyle\multiply\dimen@ by 4\divide\dimen@ by 5\fi
     \edef#2{\dimen@}}
\def\F@gin#1#2#3#4{
  \ifepsfloaded
    \checkex #1
    \iffigureexists
        \immediate\write16{(#1)}
        \begin{figure}
        \ifdim#2>\z@\reduceland{#2}{\dimen@ii}\epsfxsize=\dimen@ii\fi
        \ifdim#3>\z@\reduceland{#3}{\dimen@ii}\epsfysize=\dimen@ii\fi
        \centerline{\epsfbox{#1}}
        {#4} \end{figure}
    \else
        \immediate\write16{(#1 NOT FOUND!)}
        \begin{figure}
        \ifdim#2>\z@\reduceland{#2}{\dimen@ii}\else\dimen@ii=2in\fi
        \ifdim#3>\z@\reduceland{#3}{\dimen255}\else\dimen255=2in\fi
        \centerline{\framebox[\dimen@ii]{\rule{0pt}{\dimen255}#1}}
        {#4} \end{figure}
    \fi
  \else
    \immediate\write16{(NOT inputting #1; no epsf.tex)}
    \begin{figure}
    \centerline{\framebox[2in]{\rule{0pt}{2in}#1}}
    #4\end{figure}
  \fi}
\def\figinsertx#1#2#3{\F@gin{#1}{#2}{0pt}{#3}}
\def\figinserty#1#2#3{\F@gin{#1}{0pt}{#2}{#3}}
\def\figinsert#1#2{\F@gin{#1}{0pt}{0pt}{#2}}
\makeatother
\begin{document}
\nonsequentialeqnum 
\pubnum{94-23 \cr
 KEK-TH-410 \cr
 KEK-Preprint-94-81 \cr}
\date{August 1994}

\title{Consequence of Hawking radiation \\
from 2d dilaton black holes}

\author{Tsukasa TADA\thanks{JSPS fellow. Supported in part
by Grant-in-Aid for Scientific Research from MESC  and NSF Grant
No. PHY89-04035.}\thanks{E-mail: tada@itp.ucsb.edu}%
\thanks{Present address: Physics Department, The University of
California, Santa Barbara, California 93106}}
\address{National Laboratory for High Energy Physics (KEK)\\
Tsukuba, Ibaraki 305, Japan}
\author{Shozo UEHARA\thanks{E-mail: uehara@yukawa.kyoto-u.ac.jp}}
\address{Uji Research Center \\
	       Yukawa Institute for Theoretical Physics\\
               Kyoto University,~Uji 611,~Japan}
\maketitle

\begin{abstract}
We investigate the CGHS model through numerical calculation.
The behavior of the mass function, which we introduced in our
previous work as a ``local mass'', is examined.
We found that
the mass function takes negative values, which means that
the amount of Hawking radiation becomes greater than the initial
mass of the black hole as in the case of the RST model.
\end{abstract}

\section{Introduction}
    Two dimensional dilaton black hole models
have attracted a great deal of attention in recent years
\cite{cghs,bs,susskind,RST,bc,strominger,dealwis,hamada}.
They provide a way to treat black holes that
evaporate due to quantum effects and so may
serve as a useful toy model for real four dimensional black
holes, whose evaporation processes lead to such issues
as the well-known information problem.
Although the analysis of 2d dilaton black hole models has so far
remained within the semi-classical approximation,
this is still a great step towards the complete understanding of
the quantum behavior of black holes.

We are interested in studying the end point of these black
hole models within the semi-classical approximation.
Numerical studies of the CGHS model have been made
by several authors \cite{HaSt,lowe,PiSt},
but, little is known about the end point of 2d black holes.
In particular, it is a difficult task to estimate the total
amount of Hawking radiation produced by a black hole in
a numerical study.

In this paper we investigate 2d dilaton black hole models
with the local mass introduced in our previous
paper \cite{TaUe}.
This local mass can be related to the integral of the energy
momentum tensor in the weak field region, and hence we can
use it to calculate the total energy radiated from
the black hole.

In the next section, we review our previous work and
introduce the local mass. We will use this method to investigate
the analytical solution for evaporating black holes
in section \ref{section:RST}.
In particular, we demonstrate that the excess of
Hawking radiation over the initial incoming energy
leads to a negative value of the local mass.

We present the result of our numerical study in section
\ref{section:Numerical}.
It is shown that the CGHS model suffers from the same problem,
i.e., the excess Hawking radiation, as in the RST model.
Section \ref{section:discussion} is devoted to a discussion.

\section{Local mass}\label{section:localmass}

In this section we will explain the local mass which we
have used in \cite{TaUe} to analyze 2d quantum black hole models.
An expression of the local mass for spherically symmetric four
dimensional gravity can be found in the literature
(see {\it e.g.} \cite{fmp}).
Tomimatsu \cite{tomimatsu} used the local mass in four dimensions
to investigate an evaporating Schwartzschild black hole.
The local mass for the 2d dilaton black hole is also considered
in \cite{mgny,frolov,mann}.

The gravitational part of the action which we are going to study
consists of the 2d metric $g$, a dilaton $\phi$ and
a constant $\lambda$:
\begin{equation}
S^{(G)} = {1 \over 2 \pi}\int d^2x \sqrt{-g}
    \ e^{-2\phi}\left( R + 4 (\nabla\phi)^2+ 4 \lambda^2 \right).
\label{eqn:ichi}
\end{equation}
For the matter part, we introduce N scalar fields $f_i$:
\begin{equation}
S^{(M)}= {1 \over 2 \pi}\int d^2x \sqrt{-g} \left(
	- \frac{1}{2} \sum_{i}^{N} (\nabla f_i)^2 \right).
\end{equation}
Throughout the paper we restrict our investigation to initial
matter configurations such that a shock wave of $f_1$ comes
in at an advanced time $x^+_0$ with the total energy M.

    In the Hamiltonian formalism, we see that a certain combination
of the two Hamiltonian constraints for the gravitational part gives
the total divergence of the local mass \cite{TaUe}.
    Decomposing the two dimensional metric $g$ as
\begin{equation}
	g_{ab}=\pmatrix{ -N_0^2+{N_1^2 \over \gamma}& N_1 \cr
	N_1&\gamma\cr}~,\label{eqn:ni}
\end{equation}
one can rewrite the gravitational action in terms of
the canonical variables with the Hamiltonian constraints:
\begin{equation}
	S^{(G)}=\int \pi_\phi {\dot \phi}+ \pi_\gamma{\dot\gamma}
	-N_0 {\cal H}_0 -N_1{\cal H}_1~.\label{eqn:san}
\end{equation}
It then turns out that the following combination of
the above constraints becomes exactly the total derivative
of a scalar function:
\begin{equation}
{2\phi' \over \lambda\sqrt{\gamma}} \times {\cal H}_0
	- {4\gamma\,\pi_\gamma\,e^{2 \phi} \over \lambda}
	\times {\cal H}_1~ \equiv {\cal M}' ,\label{eqn:go}
\end{equation}
where the prime, $'$, stands for spatial derivative.
The above scalar function ${\cal M}$ can be expressed as
\begin{equation}
{\cal M} = {4\gamma\,\pi_\gamma^2 \over \lambda}\,e^{2 \phi}
	- {(\phi')^2 \over \lambda\gamma}\,e^{-2 \phi}
	+ \lambda\,e^{-2\phi}~.\label{eqn:rokub}
\end{equation}
We call the function ${\cal M}$ the local mass.

	It would be more transparent to execute the procedure
described above in the covariant formalism \cite{mann}.
The energy-momentum tensor of the gravitational action is
\begin{eqnarray}
	T_{\mu \nu}^{(G)}&\equiv &\frac{2\pi}{\sqrt{-g}}
           \frac{\delta	S^{(G)}} {\delta g^{\mu \nu}}\nonumber \\
	&=& 2 e^{-2\phi} \left[ g_{\mu\nu} \left( \nabla^2 \phi
	-(\nabla \phi)^2 + \lambda^2 \right)
	- \nabla_\mu \nabla_\mu	\phi \right].
\end{eqnarray}
If we make a current $\cal J_\mu$ from the energy-momentum tensor,
\begin{equation}
	{\cal J_\mu}\equiv T_{\mu \nu}^{(G)} \epsilon^{\nu\alpha}
	\nabla_\alpha (\frac\phi\lambda),
\end{equation}
it turns out that it is divergence-less,
$\nabla^\mu {\cal J_\mu}=0$ identically.
This is the reminiscent of what Kodama \cite{Kodama}
found in the four dimensional spherically symmetric system.
	Because of its divergence-less nature
there exists a scalar function such that ${\cal J_\mu} =
g_{\mu\nu}\epsilon^{\nu\alpha}\nabla_\alpha{\cal M}$.
In other words, the following identity holds:
\begin{equation}
	 T_{\mu \nu}^{(G)} \epsilon^{\nu\alpha}
	\nabla_\alpha (\frac\phi\lambda) \equiv
	g_{\mu\nu}\epsilon^{\nu\alpha}\nabla_\alpha
	{\cal M}. \label{eqn:TtoM}
\end{equation}
The explicit form of $\cal M$ is given by
\begin{equation}
	{\cal M}= \frac{e^{-2\phi}}{\lambda}\left[
	\lambda^2-g^{\mu\nu}\nabla_\mu \phi \nabla_\nu \phi\right]~,
	\label{eqn:covlm}
\end{equation}
which is the covariant expression of (\ref{eqn:rokub}).

There is, however, an ambiguity in the above construction,
as ${\cal M}$ may be multiplied by a constant or
have a constant added to it.
This ambiguity is fixed in the following way.
	We can relate the local mass to the flux of the energy
momentum tensor in the weak field region.
The flat space-time solution for this model is known as
the Linear Dilaton Vacuum (LDV):
\begin{eqnarray}
	&&\phi=-\frac\lambda2 (\sigma^+-\sigma^-) ,\nonumber \\
	\label{eqn:LDV}\\
	&&ds^2 =- d\sigma^+ d\sigma^- . \nonumber
\end{eqnarray}
It is easy to see that in this geometry (\ref{eqn:LDV})
the local mass (\ref{eqn:covlm}) becomes zero.
For the above geometry the identity (\ref{eqn:TtoM}) simply becomes
\begin{eqnarray*}
	&&\partial_+ {\cal M}= - ( T_{++}^{(G)}+ T_{+-}^{(G)}),\\
	&&\partial_- {\cal M}= T_{-+}^{(G)}+ T_{--}^{(G)}   .
\end{eqnarray*}
Now $T_{+-}^{(G)}$ is nothing but the gravitational part of the
equations of motion which should equal zero if quantum
effects are negligible. Hence, for the flat space-time we obtain
\begin{eqnarray}
	&& \partial_+ {\cal M}= -  T_{++}^{(G)}, \nonumber \\
	\label{eqn:flatMtoT}\\
	&&\partial_- {\cal M}=  T_{--}^{(G)}   .\nonumber
\end{eqnarray}
The ambiguity which may exist in the definition of the local mass is
fixed by the above relations (\ref{eqn:flatMtoT}) and the fact that
the local mass becomes zero for LDV (\ref{eqn:LDV}).

	If we take an asymptotically flat coordinate system, the
relations (\ref{eqn:flatMtoT}) also hold at infinity, where the
gravitational field is weak.
Assuming that the local mass becomes zero at
($x^+=0, x^-=-\infty$), integration of the first relation
in (\ref{eqn:flatMtoT}) along past null infinity ($x^-=-\infty$)
starting from $x^+=0$ gives
\begin{equation}
	{\cal M}(x^+, -\infty) = \int_0^{x^+}
	(-T_{++}^{(G)})\Big|_{x^-=-\infty}\, d\sigma^+ .
	\label{eqn:Mpastinf}
\end{equation}
In particular, the local mass at ($x^+=+\infty, x^-=-\infty$) gives
the ADM mass:
\begin{equation}
	{\cal M}(+\infty, -\infty) =
	\int_0^{\infty}(-T_{++}^{(G)})\Big|_{x^-=-\infty}
	d\sigma^+ = M_{ADM} .
\end{equation}
If we consider the integration of the second relation in
(\ref{eqn:flatMtoT}) along the $x^-$ direction at future null
infinity we obtain
\begin{eqnarray}
	{\cal M}(+\infty, x^-) &=& {\cal M}(+\infty, -\infty)
	+\int_{-\infty}^{x^-} T_{--}^{(G)}\Big|_{x^+=+\infty}\,
	d\sigma^- \nonumber\\ &=& M_{ADM} - \int_{-\infty}^{x^-}
	(-T_{--}^{(G)})\Big|_{x^+=+\infty}\,d\sigma^-~.
	\label{eqn:locmasbondi}
\end{eqnarray}
which coincides with the expression of the Bondi mass.

The shock wave of $f_1$, which represents the implosion of the
matter shell, gives the following energy momentum tensor:
\begin{eqnarray}
	T_{++}^{(M)}(\sigma^+, \sigma^-)
	&=& \frac12 \sum_i^N \partial_+ f_i \partial_+ f_i \\
	&=& M \delta(\sigma^+-\sigma^+_0) = - (T_{++}^{(G)})~,
\end{eqnarray}
where $\sigma^+_0=e^{\lambda x^+_0}$~. From (\ref{eqn:Mpastinf})
we obtain
\begin{eqnarray}
	{\cal M}(\sigma^+, \sigma^-)|_{x^- \rightarrow -\infty}
	&=& \int_0^{\sigma^+} T_{++}^{(M)} d\sigma^+  \nonumber \\
	&=& M\theta (\sigma^+-\sigma^+_0)~.
\end{eqnarray}
The above expression agrees with the result from the direct
calculation \cite{TaUe} of the local mass for the CGHS solution.

We emphasize that the local mass is the amount of
flux of the energy-momentum tensor in the weak field region.
We will see that it reflects quantum effects.
We can calculate the local mass on the matter shock-wave line
$x^+= x^+_0$ \cite{TaUe} as
\begin{equation}
	{\cal M}|_{x^+=x^+_0}=M \sqrt{\frac{-\lambda^2 x^+_0x^-}
	{-\lambda^2 x^+_0x^- - N/12}}~~.
\end{equation}
The local mass $\cal M$ diverges at the quantum singularity, where
$e^{-2\phi}$ reaches $\frac{N}{12}$.

	It is worth noting that the local mass (\ref{eqn:covlm}) can
be rewritten as
\begin{equation}
	{\cal M}= \frac{1}{4 \lambda } (e^{-2\phi}- \frac{N}{12})~R~,
	\label{eqn:MtoR}
\end{equation}
where $R$ stands for the scalar curvature of two-dimensional
geometry.
To derive the above expression (\ref{eqn:MtoR}) we have used the
equations of motion (\ref{eqn:cghs}) which include the effective term
of the quantum correction for the matter part.
Then one can compare our local mass with another mass in the
literature\cite{lowe},
\begin{equation}
	M_{eff} = \frac{1}{4 \lambda} (1- \frac{N}{12}
	e^{2\phi})^{\frac32} e^{-2 \phi} R~.
\end{equation}
Both of them coincide at future null infinity with
\begin{equation}
	m(x^-)=\lim_{x^+\rightarrow \infty}\frac{1}{4 \lambda }
	e^{-2\phi} R,
\end{equation}
which is appeared in \cite{RST}.

We have introduced a function, "local mass", which should be useful
to investigate 2d dilaton black holes.
Unfortunately, incorporating quantum effects prevents us from
obtaining an analytical solution for the system. Therefore, we shall
study it numerically.
On the other hand Russo, Susskind and Thorlacius \cite{RST} found
a model which is exactly solvable even with the quantum correction.
In the next section we analyze the behavior of the local mass
for the RST model.

\section{Behavior for RST model}\label{section:RST}
In this section we estimate the value of the local mass for the RST
model. In Ref. \cite{RST} it was demonstrated that incorporating an
(artificial) effective term
\begin{equation}
    \frac{N}{48\pi} \int d^2x\sqrt{-g}\phi R , \label{eqn:RSTterm}
\end{equation}
into the CGHS model, as well as a term from the quantum effect
of the matter, one obtains the exact solution:
\begin{eqnarray}
	&& \phi = \frac{24}{N}e^{-2\phi}
	-\frac{24 \lambda^2 x^+x^-}{N}-\frac12 \ln(\lambda^2 x^+ x^-)
	-\frac{24 M (x^+-x^+_0)}{N\lambda x^+_0}\theta(x^+-x^+_0)~,
	\nonumber \\
	&&  \\
	&&ds^2 =-e^{2\phi} dx^+ dx^- . \nonumber
\end{eqnarray}
The above solution describes the formation of a black hole and its
evaporation through Hawking radiation.

We can visualize the RST model by the behavior of the local mass at
each space time point. It is shown in Figures \ref{fig:1a} and
\ref{fig:1b}.

\figinsertx{RSTmassa.eps}{14cm}{
\caption{The behavior of the local mass for the RST model is shown in
3Dplot}\label{fig:1a}}
\figinsertx{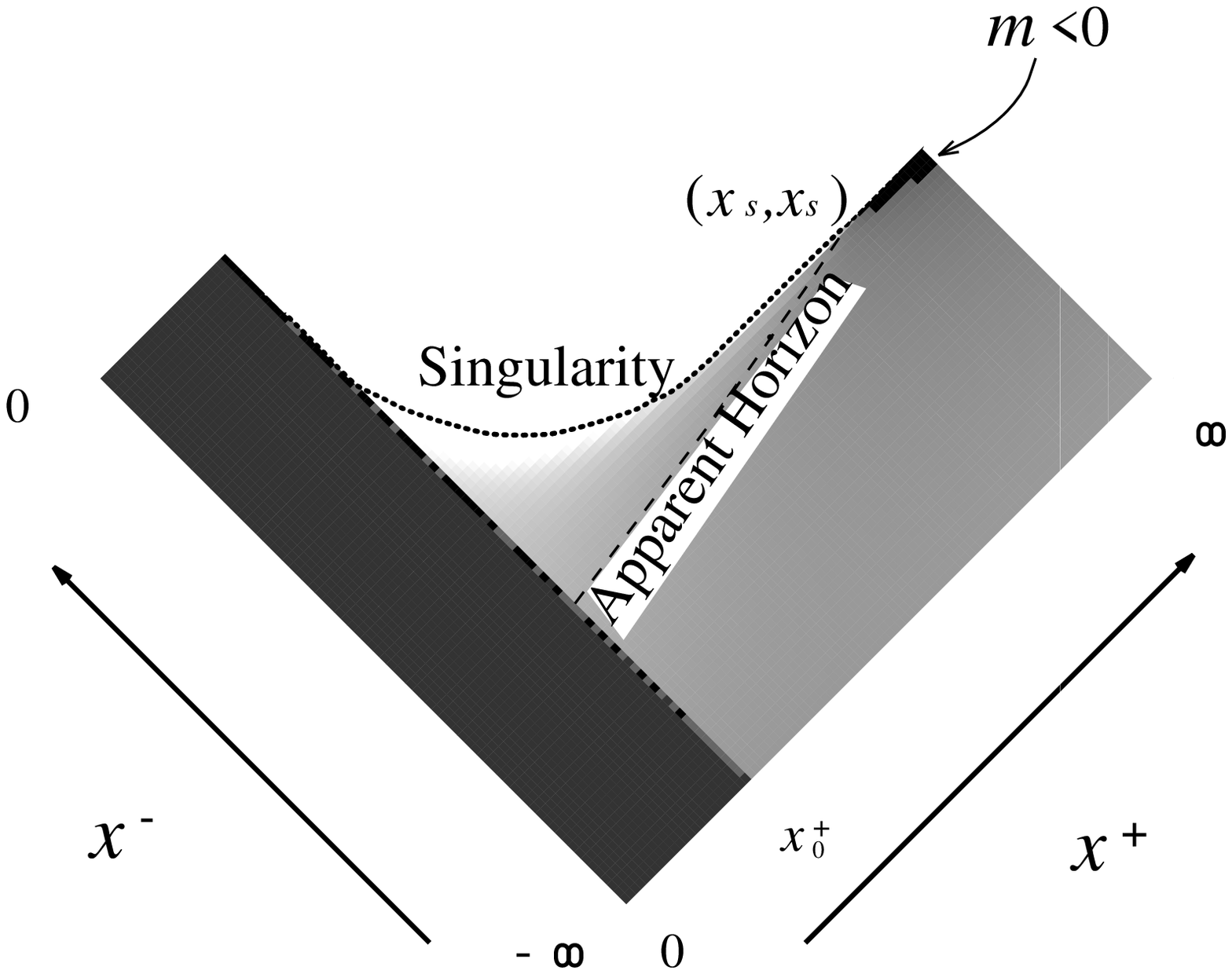}{14cm}{
\caption{The behavior of the local mass for the RST model
in density contour plot, where the density decreases
as the local mass grows.}\label{fig:1b}}

In the region $x^- \to - \infty$, the local mass shows the same
behavior as that of the original CGHS model. It is clear from the
fact that the above RST solutions reduce to the classical ones of
the CGHS model at $x^- \to - \infty$.
On the other hand, the local mass behaves differently at
$x^+=x^+_0$ as
\begin{equation}
	{\cal M}|_{x^+=x^+_0}=M \frac{-\lambda^2 x^+_0x^-}
	{-\lambda^2 x^+_0x^- - N/48}.
\end{equation}
This difference is caused by the additional effective term
(\ref{eqn:RSTterm}).

The RST model has the advantage that we can
evaluate the value of the local mass at $x^+=+\infty$ analytically.
The information at future null infinity contains that of the end
point of the black hole.
Also, at future null infinity the local mass directly
corresponds to the total energy radiated so far.
The result at future null infinity is
\begin{equation}
	{\cal M}|_{x^+\rightarrow +\infty} (x^-) = M +
	\frac{N}{48}\Bigl\{ - \frac{M}{-\lambda^2 x^+_0 x^-}
	+ \lambda\log (1-\frac{M}{-\lambda^2 x^+_0 x^-}) \Bigr\}.
\end{equation}

	The fact that in the RST model the black hole falls into
negative energy state can be reproduced exactly in our study with
the local mass.
In Ref.\cite{RST} it is claimed that up to the point where the
singularity and the apparent horizon meet together,
\begin{eqnarray}
	&&x_s^+=\frac{N\lambda x^+_0}{48M}
		(e^{48M/N\lambda}-1), \nonumber \\
	&& \nonumber \\
	&&x_s^-=-\frac{M}{\lambda^3x^+_0}\frac{1}{1-e^{-48M/N\lambda}},
\end{eqnarray}
the total amount of the energy radiated by the Hawking radiation
becomes $$M+\frac{N\lambda}{48}(1-e^{-48M/N\lambda}).$$
This exceeds the energy supplied by the matter shock wave at
$x^+=x^+_0$
by $\frac{N\lambda}{48}(1-e^{-48M/N\lambda})$.
In fact, the local mass at the future null infinity gives
\begin{equation}
	{\cal M}(+\infty,x_s^-) = -\frac{N\lambda}{48}
	(1-e^{-48M/N\lambda}).
\end{equation}
This confirms the relation between the local mass
and the energy momentum tensor (\ref{eqn:locmasbondi}).
In Figure \ref{fig:1b}  one can see the region where the local mass
becomes negative at the upper right of $(x_s^+,x_s^-)$.

\section{Numerical study of the CGHS model}\label{section:Numerical}

	The equations of motion of the CGHS model at the one-loop
level are given by
\begin{eqnarray}
&&2\left(1-{N\over 12}\,e^{2\phi}\right)\,\partial_+\partial_-\phi
	- \left(1-{N\over 24}\,e^{2\phi}\right)\,\left
	  ( 4\partial_+\phi\, \partial_-\phi +
	  \lambda^2\,e^{2\rho}\right) = 0~,\nonumber\\
	\label{eqn:cghs}\\
&&\partial_+ \partial_- \phi = \left( 1- {N\over 24}\,
	e^{2\phi} \right) \partial_+ \partial_- \rho~ , \nonumber
\end{eqnarray}
where $\rho$ stands for the so-called Liouville field:
\begin{displaymath}
	ds^2= - e^{2\rho}dx^+dx^-~.
\end{displaymath}
These equations which incorporate the quantum effects are no longer
exactly solvable and we shall solve them numerically.

	The parameter $\lambda$ can be scaled away and we put
$\lambda = 1$. The initial data surface, $x^+ = x_0^+$, is the line
of a shock-wave of an infalling matter and we will choose $x_0^+=1$.
The $\phi$ and $\rho$ on $x^+ = 1$ are those of the linear dilaton
vacuum (LDV) and those values
are given by
\begin{equation}
	\phi(1,x^-) = \rho(1,x^-) = -{1\over 2}\,\log\,(-x^-)~.
	\label{eqn:iniphi}
\end{equation}
	We shall impose boundary conditions at past null infinity
such that the solutions become the classical ones there.
Actually at a large negative value of $x^-$, we impose that
the $\phi$ and $\rho$ take the classical values.

	For fixed $x^+$ the equations (\ref{eqn:cghs}) reduce to the
ordinary differential equations for $\partial_+ \phi$ and
$\partial_+\rho$ with respect to $x^-$. We numerically integrate the
equations from a large negative value of $x^-$ through the apparent
horizon, $\partial_+ \phi = 0$, to the singularity
$e^{2\phi} = 12/N$.
	In practice we calculate up to the singularity.
With the calculated values of $\partial_+ \phi$ and $\partial_+ \rho$
we step forward in the $x^+$ direction. Actually the values of
$\partial_+\phi$ and $\partial_+ \rho$ on the shock-wave line
$x^+ = 1$ can be calculated from
Eqs.~(\ref{eqn:cghs}) \cite{susskind,TaUe},
\begin{eqnarray}
	\partial_+\phi(1,x^-) &=& -{1\over 2} +
	{M\over 2}\,\frac{1}{\sqrt{-x^-}\,\sqrt{-x^--N/12}}~,\,
	\nonumber\\
	\label{eqn:dpphio}\\
	\partial_+\rho(1,x^-) &=& -{1\over 2} - {12\,M\over N}\,
	\left(1-\sqrt{1+{N\over{12\,x^-}}}^{-1}\right)~.\nonumber
\end{eqnarray}
and they can be used to check the accuracy of the numerical
algorithm.
The numerical method which we adopted is explained in the appendix.

	The result of the singularity and the apparent horizon in the
CGHS model for $M=25$ and $N=600$ is shown in Fig.~\ref{fig:cghs-as}.
It indicates that the singularity and the apparent horizon meet
at finite $x^+$ and $x^-$ which we denote $(x^+_{EP},x^-_{EP})$.
\figinsertx{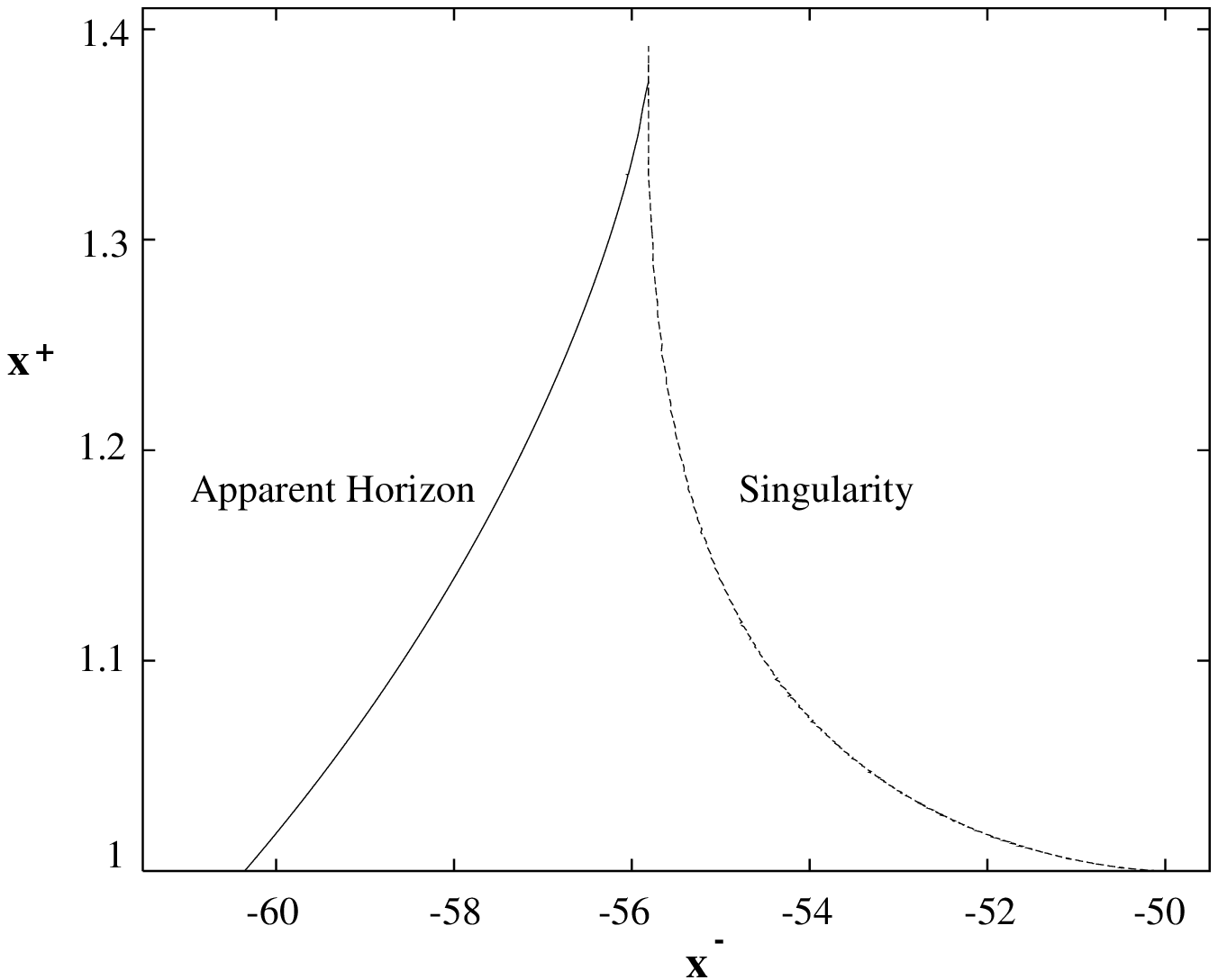}{14cm}{
        \caption{The singularity and the apparent horizon in the CGSH model.}
        \label{fig:cghs-as}
}

	The local mass function (\ref{eqn:rokub}) in the CGHS model
for the same $M$ and $N$ is shown in Fig.~\ref{fig:cghs-m} and
\ref{fig:CGHSmassb}.
\figinsertx{CGHSmassa.eps}{14cm}{
        \caption{The local mass function in the CGSH model (3Dplot).}
        \label{fig:cghs-m}
}
\figinsertx{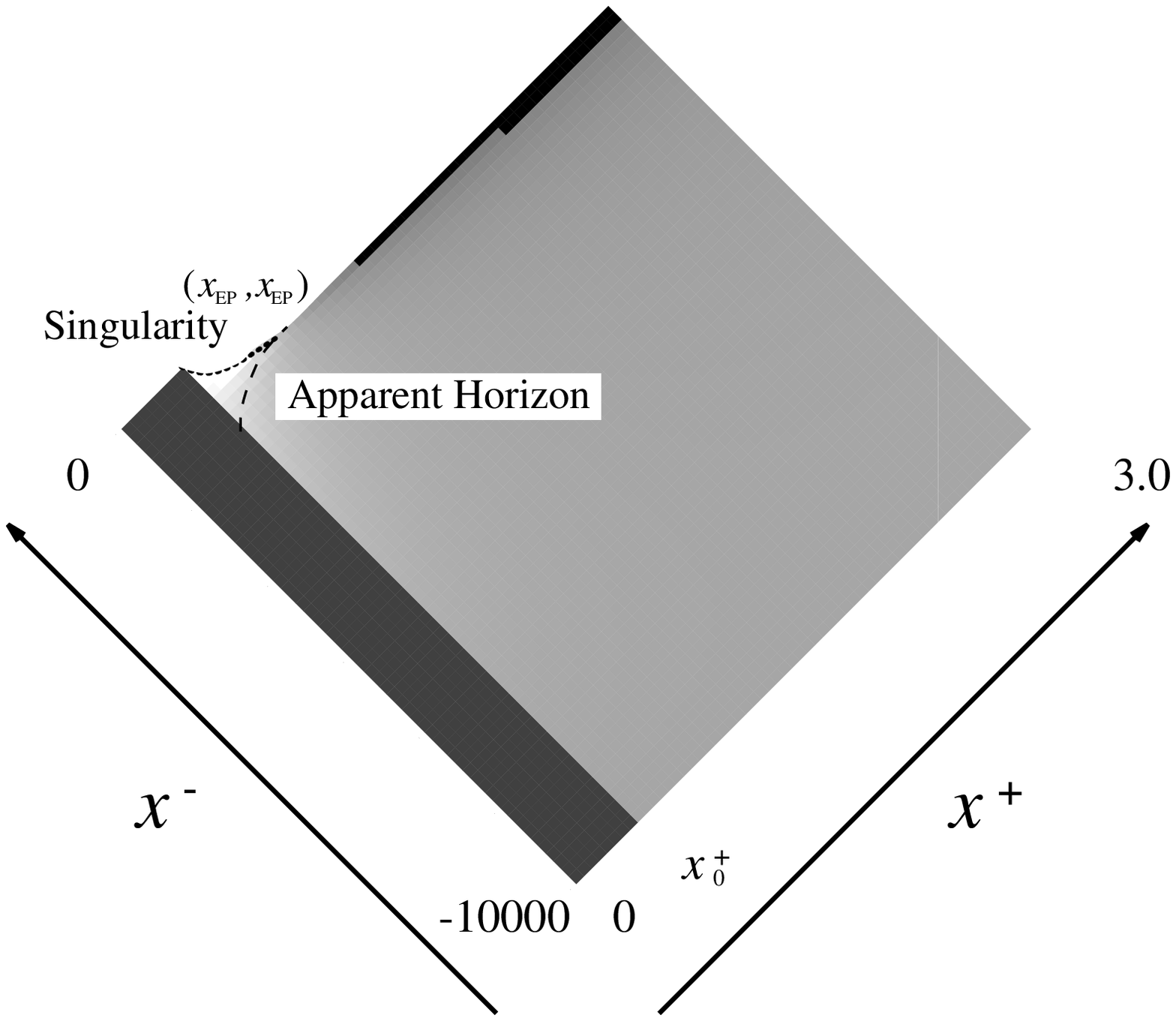}{14cm}{
        \caption{The local mass function in the CGSH model (density plot).}
        \label{fig:CGHSmassb}
}
  The figure \ref{fig:CGHSmassb} shows that the local mass becomes
negative in some region, before $x^- = x^-_{EP}$ in the $x^-$
direction and after the endpoint $x^+ > x^+_{EP}+
({\rm some\ const.})$ in the $x^+$ direction,
in a manner similar to the RST model.

\section{Discussion}\label{section:discussion}

In the previous section we found that the amount of
the Hawking radiation in the CGHS model exceeds the incoming
energy as was seen in the RST model case.
In other words, the CGHS model fails to end up as the vacuum but
evolves into a space-time whose scalar curvature is negative.

One interpretation is that this merely means the failure of
the CGHS model as a toy model for four dimensional
black hole models.
Note in particular that we have coupled scalar matter fields
to the gravitational field in a way that is different from
the four dimensional case.
This coupling of the matter fields enables us to
calculate the back reaction but could also cause the excess of the
radiation energy.

One possible modification of the CGHS model may come from
string theory.
There have been many attempts to resolve the problems
associated with black holes by applying string theory
and one approach along this line is the following.
The term $(1/4) e^{-2\phi}\alpha ' R_{abcd}R^{abcd}$\cite{CMP},
where $\alpha '$ is the inverse of the string tension,
will provide string corrections to the model.
It would be interesting to estimate the effect of this using
present local mass analysis.

For further investigation it may also be interesting to analyze
a black hole model proposed by Nojiri and Oda \cite{NojiriOda}.
The model is described by SL(2,R)/U(1) gauged
WZW model deformed by (1,1) operator and they
claim that the black hole reaches a zero mass state and
evaporates completely.

\vspace{1cm}
\noindent{\bf Acknowledgment}
We would like to thank T. Eguchi for suggestive discussion.
One of the authors (T.T) would like to thank Soryuushi Shougakukai
for the financial support at the early stage of the present work.
His thanks are also due to ITP members for their hospitality
and the participants of the conference `Quantum Aspects of Black
Holes', especially to A. Strominger for discussions.
Finally useful comments by D. Marolf to the manuscript is gratefully
acknowledged.

\begin{appendix}
\def\theequation{\Alph{section}.\arabic{equation}}
\section{The Numerical Method}
	The two dimensional surface $(x^+,x^-)$ is represented by a
two dimensional lattice with equal spacing in each direction. The
lattice spacing in the $x^+$ direction is much narrower than
that in the $x^-$ direction.

The equation (\ref{eqn:cghs}) can be symbolically represented by
\begin{eqnarray}
	\partial_-\partial_+\phi &=&
	f(\phi,\rho,\partial_+\phi)~,\nonumber\\
	\label{eqn:app1}\\
	\partial_-\partial_+\rho &=&
	g(\phi)\,\partial_-\partial_+\phi~.\nonumber
\end{eqnarray}
and for fixed $x^+$ the first equation can be seen as an ordinary
differential equation of the first order for $\partial_+\phi$.
We use the fourth-order Runge-Kutta method to get $\partial_+\phi$ in
the $x^-$ direction.
We have adopted the ``Cubic Spline Interpolation'' method to get
values of an arbitrary $x^-$ at fixed $x^+$ and $\partial_-\phi$ and
$\partial_-\rho$ are computed with a fourth-order backward
differential method.

	The values of $\phi$ and $\rho$ on the next $x^+ = {\rm
const.}$ line are obtained from $\partial_+\phi$ and
$\partial_+\rho$ by linear extrapolation.

\end{appendix}


\begin{thebibliography}{99}
\bibitem{cghs}{C.G.~Callan, S.B.~Giddings, J.A.~Harvey and
	A.~Strominger, Phys. Rev. {\bf D45} (1992) R1005.}
\bibitem{bs}{T.~Banks, A.~Dabholkar, M.R.~Douglas and
	M.~O'Loughlin, Phys. Rev. {\bf D45} (1992) 3607.}
\bibitem{susskind}{J.G.~Russo. L.~Susskind and L.~Thorlacius,
Phys. Lett. {\bf B292} (1992) 13}
\bibitem{RST}{J.G.~Russo. L.~Susskind and L.~Thorlacius,
	Phys. Rev. {\bf D46} (1992) 3444;
{\it ibid.} {\bf D47} (1993) 533.}
\bibitem{bc}{A.~Biral and C.~Callan,
		Nucl. Phys. {\bf B394} (1993) 73.}
\bibitem{strominger}{A.~Strominger,
           Phys. Rev. {\bf D46} (1992) 4396.}
\bibitem{dealwis}{S.P.~de~Alwis,
      Phys. Lett. {\bf B289} (1992) 278;
{\it ibid.} {\bf B300}(1993) 330.}
\bibitem{hamada}{K.~Hamada, Phys. Lett. {\bf B300} (1993) 322.}
\bibitem{HaSt}
{ S.W. Hawking and J. M. Stewart, Nucl. Phys. {\bf B400} (1993) 393.}
\bibitem{lowe}
{ David A. Lowe, Phys. Rev. {\bf D47}(1993) 2446.}
\bibitem{PiSt}
{ Tsvi Piran and Andrew Strominger,
Phys. Rev. {\bf D48} (1993) 4729.}
\bibitem{TaUe}{T. Tada and S. Uehara, Phys. Lett. {\bf B305} (1993)
23}
\bibitem{fmp}{W.~Fischler, D.~Morgan and J.~Polchinski, Phys. Rev.
	{\bf D42} (1990) 4042.}
\bibitem{tomimatsu}{A.~Tomimatsu, Phys. Lett. {\bf B289} (1992) 283.}
\bibitem{mgny}{ M. D.~McGuigan, C.R.~Nappi and S. A.~Yost,
         Nucl. Phys. {\bf B375} (1992) 421}
\bibitem{frolov}{ V.~P.~Frolov, Phys. Rev. {\bf D46} (1992) 5383}
\bibitem{mann}{R.~Mann,  Phys. Rev. {\bf D47} (1993) 4438.}
\bibitem{Kodama}{H. Kodama, Prog. Theor. Phys. {\bf 63} (1980) 1217.}
\bibitem{CMP}{ C. G. Callan, R. C. Myers and M. J. Perry, Nucl. Phys.
{\bf 311} (1988/89) 673.}
\bibitem{NojiriOda}{S.~Nojiri and I.~Oda, preprint NDA-FP-10-93,
hepth@xxx/9302024 (1993); Phys. Rev. {\bf D49} (1994) 4066.}
\end{thebibliography}
\end{document}